\documentclass[aps,prl,reprint]{revtex4-2}
\usepackage{graphicx}% Include figure files
\usepackage[dvipsnames, svgnames, x11names]{xcolor}  % 一般放得靠前

\usepackage{url}
\usepackage{amsmath}
\usepackage[caption=false,farskip=0pt,labelfont={bf}]{subfig}
\usepackage{dcolumn}% Align table columns on the decimal point
\usepackage{bm}% bold math
\usepackage{color}
\usepackage{booktabs} %需要加载宏包{booktabs}
\usepackage{mathrsfs}

\usepackage[pagewise]{lineno}
\begin{document}

\title{Clusters and collective motions in Brownian vibrators}

\author{Yangrui Chen$^1$ and Jie Zhang$^{1,2}$}
\email{jiezhang2012@sjtu.edu.cn}
\affiliation{$^1$School of Physics and Astronomy, Shanghai Jiao Tong University, Shanghai 200240, China}
\affiliation{$^2$Institute of Natural Sciences, Shanghai Jiao Tong University, Shanghai 200240, China}
		
\date{\today}

\begin{abstract}
Using Brownian vibrators, where single particles can undergo Brownian motion under vibration, we experimentally investigated self-organized structures and dynamics of quasi-two-dimensional (quasi-2d) granular materials with volume fractions $0.111\le\phi\le0.832$. 
We show rich structures and dynamics in hard-disk systems of inelastic particle collisions, with four phases corresponding to cluster fluid, collective fluid, poly-crystal, and crystal.  While poly-crystal and crystal are strikingly similar to the equilibrium hard disks, the first two phases differ substantially from the equilibrium ones and the previous quasi-2d experiments of uniformly driven spheres. Our investigation provides single-particle-scale evidence that granular materials subject to uniform random forcing are weakly cohesive with complex internal structures and dynamics. Moreover, our experiment shows that large-scale collective motion can arise in a purely repulsive hard-disk system. The collective motion emerges near $\phi=0.317$, where the most significant clusters span half of the system, and disappears near $\phi=0.713$, around which the system crystallizes and the melting transition occurs in the equilibrium hard disks.
\end{abstract}

\maketitle

\paragraph{Introduction.} 
Collective motion, where microscopic components exhibit large-scale correlated activities in space and time, is ubiquitous in soft \cite{Nagel-RevModPhys.89.025002,chaikin_lubensky_1995}  and active matter\cite{shankar2022topological,shaebani2020computational}.
The collective motion in the active matter is well understood. The flocks of birds can be modeled as flying spins whose moving directions are critical to the neighboring alignment \cite{vicsek1995novel, Toner-Tu-PhysRevLett.75.4326, Toner-Tu-PhysRevE.58.4828}. Rod-like particles can be described as active nematics by considering local alignment and volume exclusion \cite{doostmohammadi2018active}. When particles are self-spinning, a remarkable topologically protected edge mode can occur due to the nonreciprocal interactions \cite{banerjee2017odd}. \par

Collective motion in granular materials often relates to jamming or a particle's polarity and shape. In quasi-statically sheared dense granular packings, contacts and contact forces are essential in floppy modes\cite{liu2010jamming, van2009jamming}, plastic deformation\cite{maloney2006amorphous,wang2020connecting}, and turbulent-like vortices\cite{Farhang-PhysRevLett.89.064302, sun2022turbulent}.  Under vibration, inelastic collisions are prominent, where the particle's polarity and shape are essential. Self-propelling polar particles\cite{deseigne2010collective, deseigne2012vibrated, kudrolli2008swarming} and self-spinning disks\cite{scholz2018rotating,liu2020oscillating} are active matter in disguise, exhibiting collective behaviors. Rod-like particles can be described in active nematics \cite{kumar2014flocking, kudrolli2008swarming, narayan2010phase,kumar2011symmetry, doostmohammadi2018active}. 
Simulations \cite{fily2012athermal} found that the alignment interaction is crucial in the flocking of self-propelling disks, the absence of which can only induce motility-induced phase separation, as seen in active colloidal experiments \cite{redner2013structure, buttinoni2013dynamical}. For spheres or disks with no preferred translational or rotational directions, under 2d uniform driving, no collective motions are expected, as corroborated by the previous experiments\cite{olafsen1998clustering, Shattuck-PhysRevLett.98.188301,losert1999velocity, tatsumi2009experimental,melby2005dynamics}.

In this letter, we systematically investigate self-organized structures and collective dynamics using Brownian vibrators -- disks with alternating inclined legs under the rim -- in quasi-2d experiments with a wide range of packing fractions $\phi$. Rich structures and dynamics emerge due to inelastic particle collisions, corresponding to cluster fluid, collective fluid, poly-crystal, and crystal.  While poly-crystal and crystal are strikingly similar to the equilibrium hard disks\cite{Mitus-PRE-1997, alder1957phase, Zollweg-PRB, Lee-PRB, Weber-EPL, Alonso-PRL}, the first two phases differ substantially from the equilibrium ones and the previous quasi-2d experiments of uniformly driven spheres \cite{olafsen1998clustering, Shattuck-PhysRevLett.98.188301, Olafsen-Urbach-2005, Pacheco-Vazquez-2009, Aranson-PRL-2000, Howell-PRE-2001, Oyarte-PRE-2013, Rivas-2011a, Rivas-2011b, Rivas-2012, Roeller-PRL-2011, Nahmad-PRE, Neel-PRE-2004, Clewett-2012, Luu-2013}. Our investigation shows that granular materials subject to uniform random forcing are weakly cohesive with complex internal structures and dynamics and that large-scale collective motion can arise in a purely repulsive hard-disk system.\par

In cluster fluid ($\phi\le0.270$), particle clusters are of power-law size distributions with an exponential cutoff. At $\phi=0.317$, where the large-scale collective motion initiates, the cluster size distribution develops a flat fat tail, and correspondingly, large clusters appear and span half of the system’s area.  The collective motion terminates near $\phi=0.713$, where crystallization begins and whose value is close to the melting transition in equilibrium hard disks. Structural relaxations differ significantly over different phases and length scales: the cluster fluid relaxes almost like a liquid but with weak subdiffusion over large scales, reflecting the fractal nature of cluster fluid; the collective fluid shows superdiffusion at large scales and progressively subdiffusive glassy behaviors at small scales with $\phi$; the relaxation of poly-crystals is related to the grain boundary particles.

\begin{figure}[htpb]
\centering

\includegraphics[width=8.6cm]{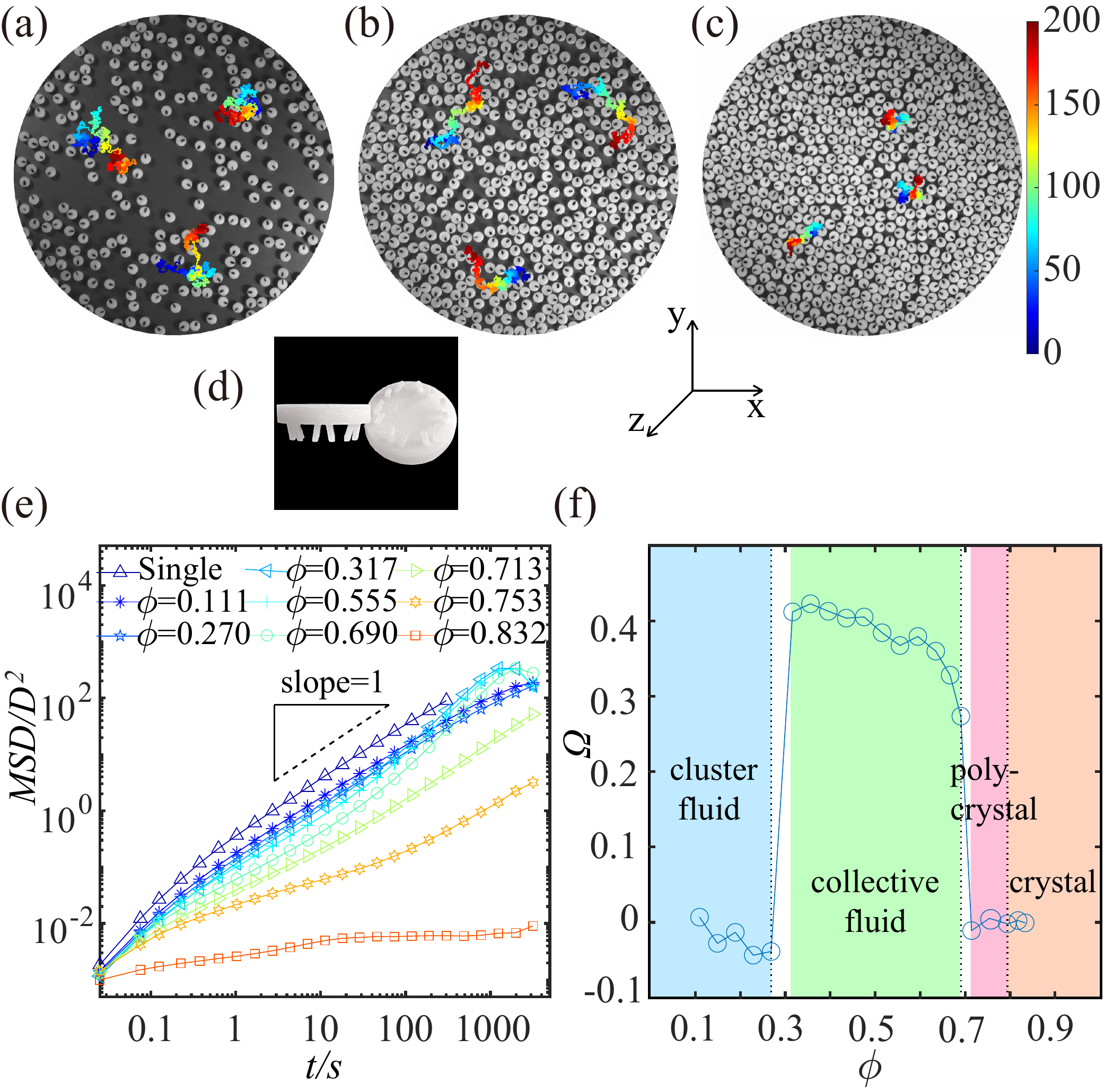}

\caption{\label{fig1} (a-c) Snapshots of particle configurations at $\phi=0.270,0.555,0.713$, respectively. The color bar represents the selected particles’ positions in 0-200s. (d) The close look of a particle whose diameter is $D=16mm$, thickness is $3mm$, and leg’s height is $3mm$. (e) Mean Square Displacement of the particle’s translational motion. (f) Granular cluster fluid ($\phi\le 0.270$), collective fluid ($0.317 \le \phi\le\ 0.690$), poly-crystal ($0.713 \le \phi\le\ 0.753$), and crystal ($\phi \ge 0.793$). $\Omega$(blue circle) is the average curl of the particle displacement field within $\Delta t=100s$.}
\end{figure}

\paragraph{Experimental setup.} 
Our experimental setup mainly consists of a horizontal confined layer of quasi-2d monodisperse Brownian vibrators placed on top of an aluminum plate driven vertically along the z-axis by an electromagnetic shaker. A flower-shaped boundary confines particles and prevents creep particle motion along the boundary\cite{chen2022high}. We discard particles within three layers next to the boundary to reduce boundary effects. Snapshots of particle configurations are shown in Fig.~\ref{fig1}(a-c), corresponding to the cluster fluid (a), collective fluid (b), and poly-crystal (c). 
An individual particle is Brownian-like with uncorrelated translation and rotation with Gaussian distributions of zero means\cite{chen2022high}. A Brownian vibrator is disk-shaped with 12 alternately inclined supporting legs, as shown in Fig.~\ref{fig1}(d). The legs are bent inward by $18.4^\circ$ and alternately deviate from the mid-axis plane by $\pm 38.5^\circ$ to randomize horizontal motion. More details can be found in Ref.\cite{chen2022high}.
\par

For a given $\phi$, we randomly place a certain number of particles on the aluminum plate and run for two hours to obtain an initial state. The vibration frequency $f =100Hz$, and the maximum acceleration $a=3g$ with $g=9.8 m/s$. The amplitude $A\equiv a/(2\pi f)^2= 0.074mm$, yielding a negligible particle’s vertical displacement. We capture particle configurations with a CCD camera at 40 frames/s for an hour for further processing.\par

\paragraph{Dynamics.} 
In Figs.~\ref{fig1} (a-c), to avoid overclouding, we only draw the trajectories of three particles to illustrate different dynamics of the clustered fluid (a), collective fluid (b), and poly-crystal (c).  At $\phi=0.270$, particles move randomly at all times in Fig.~\ref{fig1} (a).   At $\phi=0.555$, particles move randomly for $\Delta t<10s$, and however, they move collectively for $\Delta t>100s$ in Fig.~\ref{fig1} (b). At $\phi=0.713$, particles diffuse around slowly in Fig.~\ref{fig1} (c). \par

The translational mean square displacements (MSD) are shown in Fig.~\ref{fig1} (e). When $\phi \le 0.270$, particles move quasi-ballistically for $t\lesssim0.1s$ before diffusing. The slope decreases slightly below one for $t\gtrsim100s$.  When $\phi=0.317,0.555,0.690$ in the collective fluid phase, particles move quasi-ballistically for $t\lesssim0.1 s$ before sub-diffusing for $0.1s\lesssim t\lesssim 20s$. However, when $t\gtrsim20s$, the slope is above one, showing super-diffusive behaviors corresponding to the large-scale collective motion. Moreover, the MSD peaks around the 2000s manifest the global collective motion. The MSD of $\phi=0.713$ is divided into three parts: the quasi-ballistic motion for $t \lesssim 0.1s$, the sub-diffusion for $0.1 \lesssim t \lesssim 30s$, and the diffusion for $t\gtrsim 30s$ due to particles at grain boundaries (See Figs. S(1-2) of the Supplementary Materials(S.M.)\cite{SM-Chen}). As $\phi$ increases to $\phi=0.753$, the MSD slope gradually decreases and eventually down to 0 till $\phi=0.793$, beyond which e.g., at $\phi=0.832$, only a single crystal exists. We characterize the large-scale collective motion using the nonzero vorticity $\Omega$ -- the average curl of the particle displacement field, as shown in Fig.~\ref{fig1} (f). The computation details of $\Omega$ using the particle's displacement fields can be seen in the S.M.\cite{SM-Chen}. The crystallization above $\phi=0.713$ in Fig.~\ref{fig1} (f) is similar to the early experiment \cite{Shattuck-PhysRevLett.98.188301}, where spheres were sandwiched and vertically vibrated between two horizontal plates, and when $0.652< \phi< 0.719$,  their system shows sub-diffusive, caging-type behaviors on MSD at intermediate time scales, similar to the curve of $\phi=0.690$ in Fig.~\ref{fig1} (f). However, within $0.652< \phi< 0.719$, no large-scale motions are observed in Ref.\cite{Shattuck-PhysRevLett.98.188301}; a so-called “isotropic fluid phase” was observed for $\phi<0.652$ \cite{ Shattuck-PhysRevLett.98.188301, shattuck-PRL-crystal}. Moreover, our system is locally more ordered as shown in Figs. S(1-2) of the S.M.\cite{SM-Chen}.
Interestingly, $\phi=0.713$ of poly-crystal in Fig.\ref{fig1} (f) is nearly identical to the melting-transition point $\phi_s=0.716$ predicted for the equilibrium hard disks \cite{Mitus-PRE-1997}. Furthermore, $\phi=0.690$ of collective fluid in Fig.\ref{fig1} (f) is identical to the value of the pure fluid $\phi_f$ \cite{Mitus-PRE-1997}.  Note that the precise values of $\phi_s$ and $\phi_f$ may vary slightly depending on the simulation methods \cite{Mitus-PRE-1997, alder1957phase, Zollweg-PRB, Lee-PRB, Weber-EPL, Alonso-PRL}. We do not have data points within $0.690<\phi<0.713$ due to the discrete increment of $\phi$ in our experiment. However, particle configurations at $\phi=0.690$ and $0.713$ show different symmetries in Fig. S4 of the S.M.\cite{SM-Chen}.

\begin{figure}[htpb]
\centering
\includegraphics[width=8.6cm]{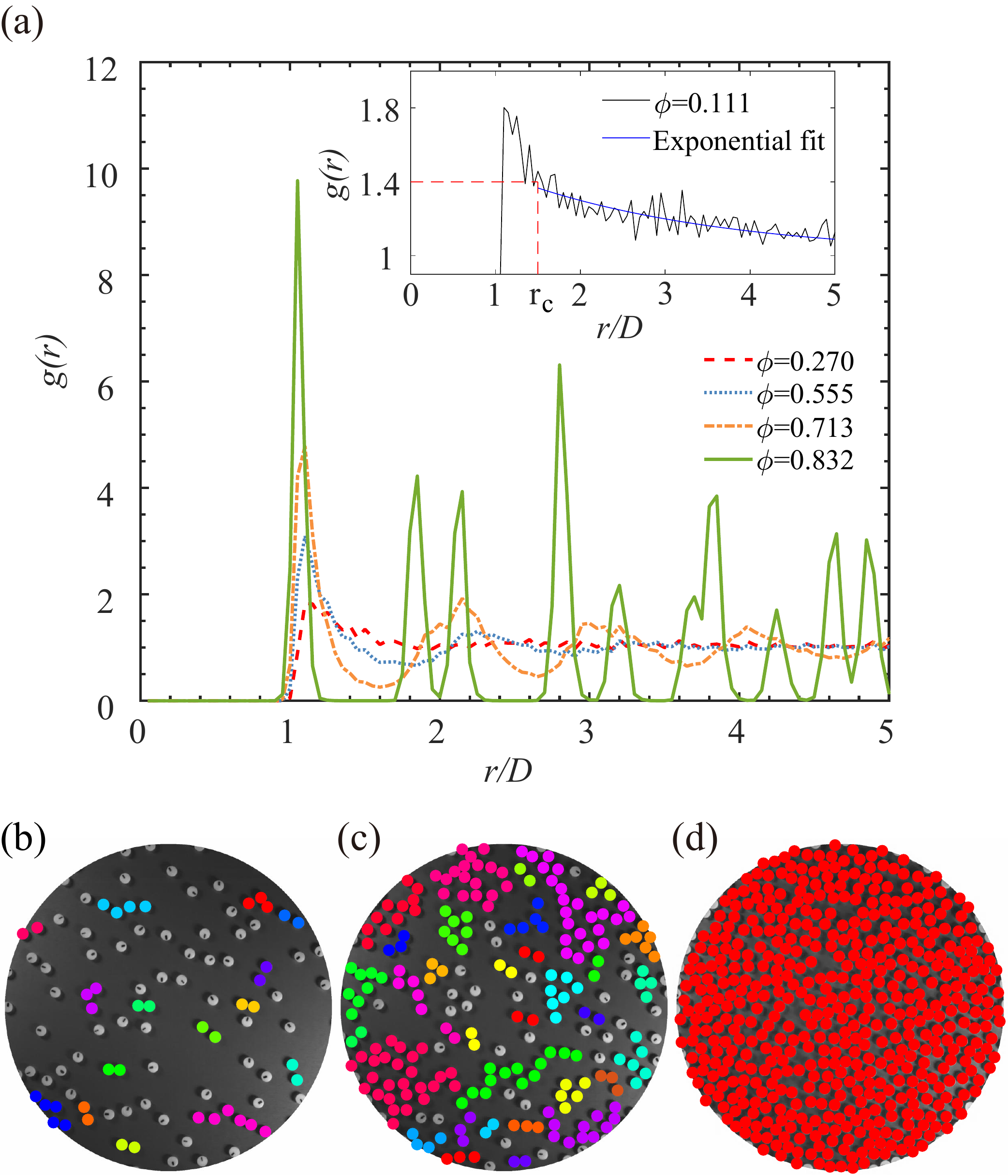}
\caption{\label{fig2} (a) Pair correlation functions $g(r)$. The red dashed line in the inset draws the half-peak height at $r_c$= 1.5D. 
(b-d) Snapshot of particle configurations at $\phi=0.111,0.270,0.555$, respectively, with clusters painted in color. The uncolored particles are single-particle clusters.
}
\end{figure}

\paragraph{Structure.} 
Fig.~\ref{fig2} (a) plots pair correlation functions $g(r)$ of different dynamical phases at given $\phi’s$. The first peak of $g(r)$ describes the mean distance $d_1$ between neighboring particles. At $\phi=0.832$, the sharp peaks at $\sqrt{3}d_1$, $2d_1$ and $ \sqrt{7}d_1$ indicate an almost perfect crystal of hard disks. At $\phi =0.713$, the system forms a poly-crystal, showing a slightly larger value of $d_1$ on $g(r)$ than that of $\phi=0.832$. Moreover, the peak at $\sqrt{3}d_1$ symbol-izes the triangular lattice. Still, it is almost buried within an extensive shoulder of the peak around $2d_1$ due to grain boundaries.  At $\phi =0.555$, the system forms a collective fluid, where the first peak of $g(r)$ shifts further to the right than that of $\phi =0.713$. The crystalline feature is washed out as the double peaks near $\sqrt{3}d_1\sim 2d_1$ disappear entirely and are replaced with a broad single peak, indicating liquid-like structures. Moreover, there are no visible peaks after the second peak in contrast to the poly-crystal. Surprisingly, in $g(r)$ of $\phi=0.270$, although the second peak disappears, the first peak survives, which indicates that two neighboring particles tend to stay close within a certain distance, tending to form chain-like clusters. In Fig. S8 (a) of the S.M., we compare the $g(r=D)$ versus $\phi$ between our system and equilibrium hard disks and the early experiments\cite{shattuck-PRL-crystal}. In the inset of Fig.~\ref{fig2} (a), we plot the $g(r)$ at $\phi=0.111$, from which we define the threshold $r_c$, corresponding to the half-height of the first peak. The blue solid line shows that $g(r)$ decays exponentially for $r>r_c$, whereas it decays more rapidly for $r\le r_c$ and the peak is much higher than that of the equilibrium hard disks\cite{chae1969radial} (See also Fig. S8 (b) of the S.M. for a quantitative comparison). We use $r_c$ to define and identify clusters (See more details in the S.M.\cite{SM-Chen}). \par

The clusters are drawn in colors in Fig.~\ref{fig2} (b-d) for particle configurations of $\phi=0.111, 0.270, 0.555$, respectively. At $\phi=0.111$, there are many single-particle clusters and chain-like clusters. At $\phi=0.270$, the average cluster size increases with more complex cluster shapes, showing the diffusion-limited-aggregation characteristics of cluster formation\cite{Witten-dla}. At $\phi=0.555$ in Fig.~\ref{fig2} (d), there is a single giant cluster in the system. Note that these clusters differ from those in literature\cite{goldhirsch1993clustering, esipov1997granular,olafsen1998clustering,caprini2020spontaneous, Roeller-PRL-2011, Nahmad-PRE, Neel-PRE-2004, Clewett-2012, Luu-2013}, where cluster particles are in close contact, forming a highly compact solid. Fig.~\ref{fig2} provides microscopic physical evidence that dry granular materials subjected to random forcing form low-surface-tension fluids at sufficiently low densities, which is consistent with the early experiments of the spinodal phase separation \cite{Clewett-2012} and the capillary-like interface fluctuations \cite{ Luu-2013} in cohesionless granular systems.
\par

\begin{figure}[htpb]
\centering
\includegraphics[width=8.6cm]{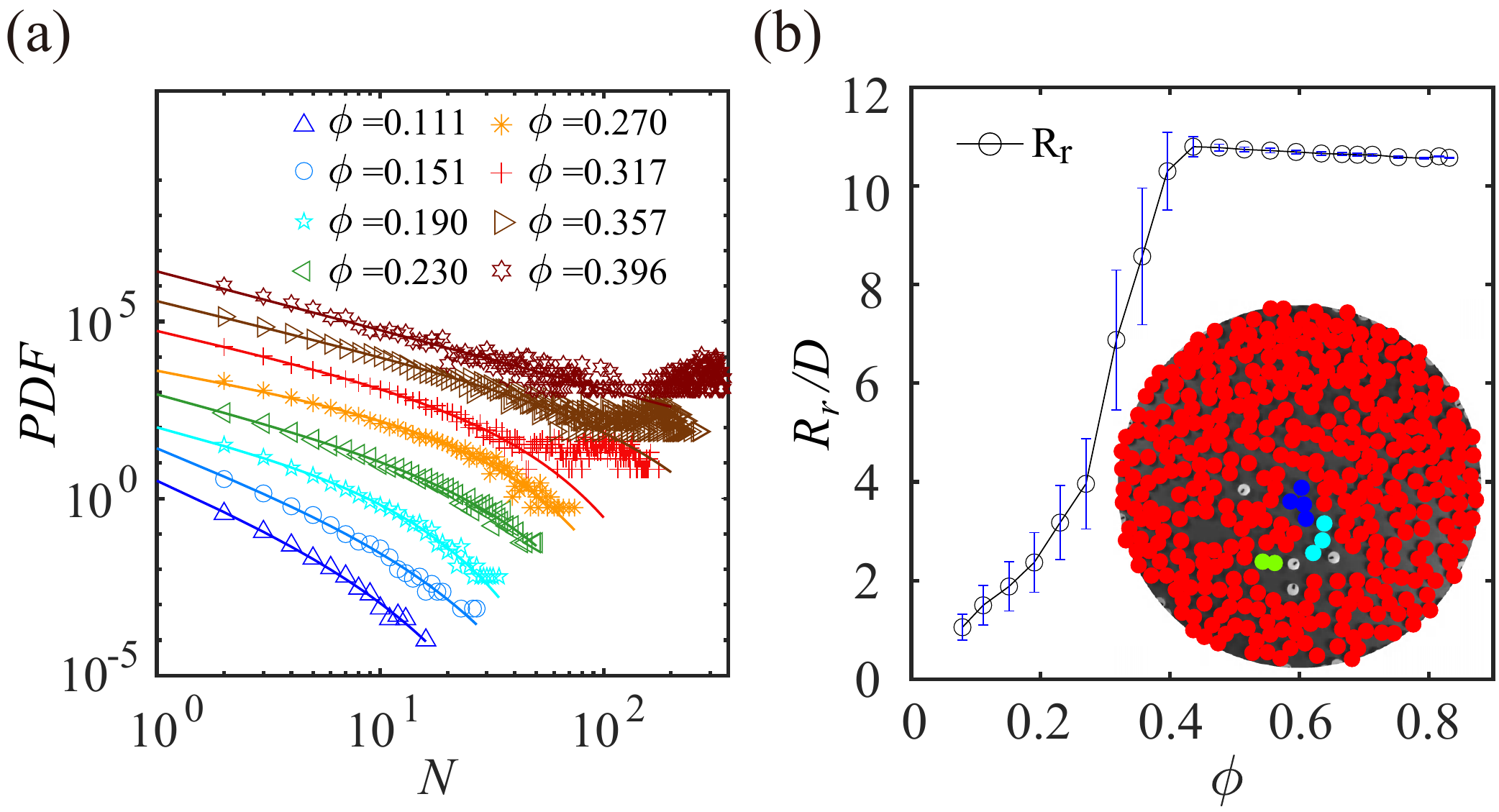}
\caption{\label{fig3} (a) Distributions of the cluster size $N$. Curves are consecutively shifted upwards by $10$ to enhance visibility. The solid line is the fit of $P \propto N^{a}e^{-N/b}$. (b) The cluster radius of gyration $R_r$ versus $\phi$. The inset shows the particle configurations at $\phi=0.436$ with clusters painted in color. }

\end{figure}

\begin{figure}[htpb]
\centering
\includegraphics[width=8.6cm]{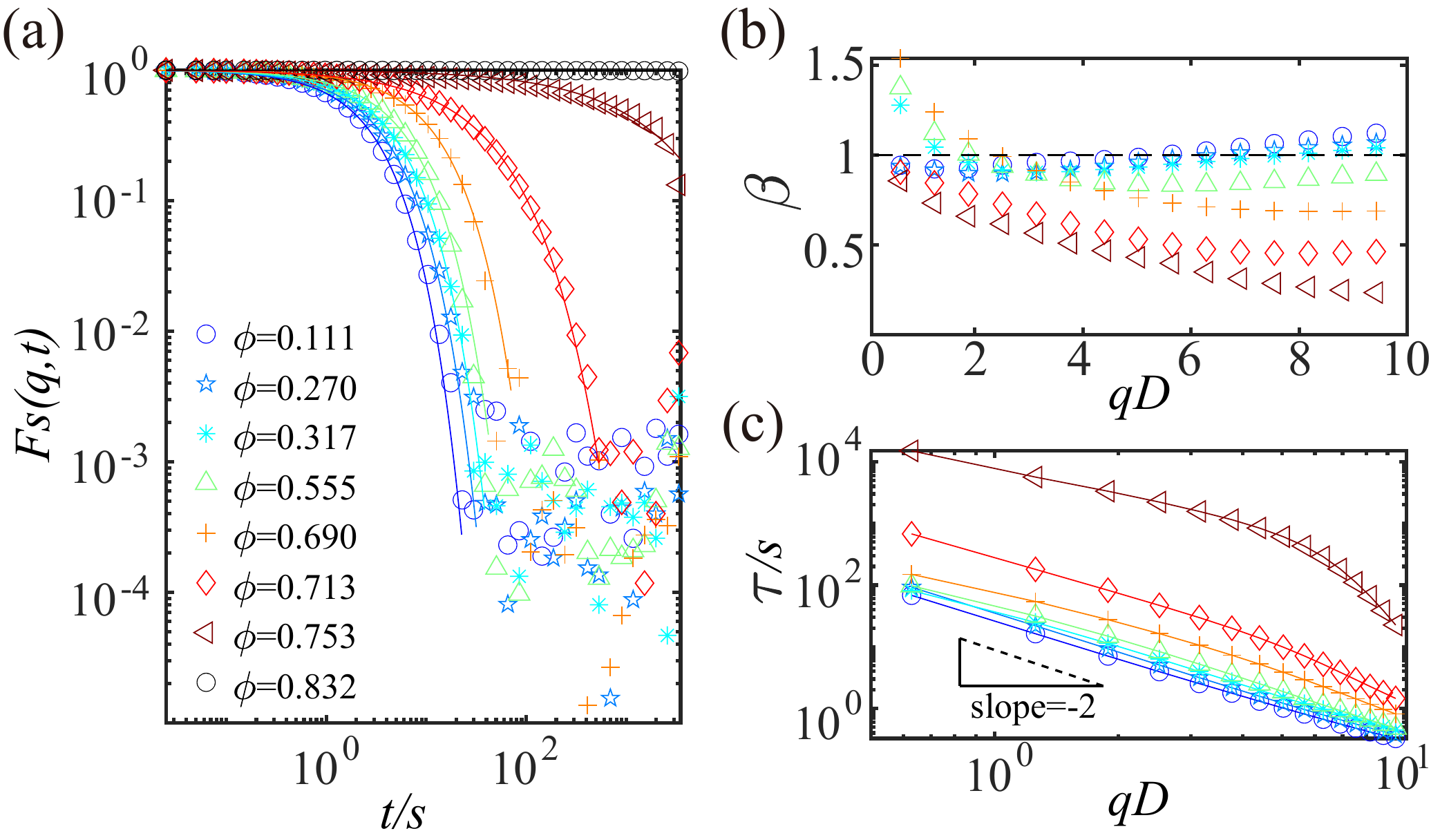}
\caption{\label{fig4} (a) Self intermediate scattering functions $F_s(q,t)$ with $q=\pi/D$. The solid lines are the stretched exponential fitting $F_s(q,t)= exp(-(t/\tau)^{\beta})$ above the noise level 0.006. (b-c) The parameters $\beta$ and $\tau$ versus q.}
\end{figure}

Fig.~\ref{fig3} (a) plots the distributions of cluster size, i.e., the number of particles of a cluster averaged over all configurations of all time. Fig.~\ref{fig3} shows that the PDFs in the cluster fluid conform to power-law distributions with exponential cutoffs, i.e., $PDF \propto N^{a}e^{-N/b}$. The fitting parameters $a$ and $b$ versus $\phi$ are shown in Fig. S9 of the S.M.\cite{SM-Chen}, where $a$ starts decreasing while $b$ increases rapidly around $\phi=0.317$.  In contrast, the PDFs in the collective fluid show an additional flat fat tail, suggesting the existence of a giant cluster comparable to the system size. Fig.~\ref{fig3} (b) plots the radius of gyration $R_r$ of the largest cluster versus $\phi$:  $R_r$ increases rapidly around $\phi=0.317$, with the crescent or quarter moon-shaped largest cluster spanning half of the system and fluctuating in time; at $\phi=0.436$, a typical largest cluster is shown in the inset, with a pocket of small clusters to cause a slight overshoot of $R_c$ that gradually reaches a plateau for $\phi>0.436$.
\par

We next use the intermediate scattering function $F_s(q,t)$ to characterize the structural relaxation for a given wave vector $\vec q$:
\begin{align}
    F_s(q,t) &= \frac{1}{N}\Sigma_{j=1}\langle e^{-i\vec q\cdot (\vec r_j(t)- \vec r_j(0))} \rangle  \nonumber \\
            &=\frac{1}{N}\Sigma_{j=1}e^{-\frac{1}{4}q^2\langle (\vec r_j(t)- \vec r_j(0))^2 \rangle},
\end{align}
where $r_j(t)$ refers to the trajectory of the particle $j$ at time $t$, $N$ is the total particle number, and the average $\langle\rangle$ is over all possible initial configurations $\vec r_j(0)$. Note that the second equality follows from assuming that the random variable $\vec r_j(t)- \vec r_j(0)$ obeys a Gaussian distribution\cite{binder2011glassy}, which is approximately valid to a certain extent, as shown in Figs. S(5-7) in the S.M.\cite{SM-Chen}. 
\par

Fig.~\ref{fig4} (a) shows the $F_s(q,t)$ with $q=\pi/D$, corresponding to the scale of $2D$. $F_s(q,t)$ decay monotonically with an increasing relaxation time with $\phi$, except at $\phi=0.832$, where the system forms a single crystal. We fit $F_s(q,t)$ with an stretched exponential function $F_s(q,t)= exp(-(t/\tau)^{\beta})$ and plot $\beta$ and $\tau$ for different $q$ in Fig.~\ref{fig4} (b-c), respectively. There are three branches of $\beta(q)$ in Fig. ~\ref{fig4} (b), corresponding to the cluster fluid, collective fluid, and poly-crystal. 
When $\phi=0.111$ and $\phi=0.270$, Fig. ~\ref{fig4} (b) shows that $\beta <1$ for small $q$ and increase above one at $q\approx 2\pi/D$, corresponding to the particle size $D$.  The weak subdiffusion at large scales is consistent with Fig.~\ref{fig1} (f), reflecting the fractal nature of clusters. 
The collective fluid shows more complex $\phi$ dependent behaviors of $\beta(q)$.  For $q<2/D$, corresponding to the scales larger than $\pi D$, $\beta(q)>1$ for all $\phi$, showing characteristics of the large-scale collective motions.  For $q\ge2/D$, when $\phi=0.317$, $\beta$ stays slightly below one within $2/D<q<2\pi/D$ and then goes above one for $q>2\pi/D$, similar to the curves of the cluster fluid. When $\phi=0.555$ and $\phi=0.690$, $\beta$ remain below one for $q>2\pi/D$ and reach a value around 0.7 in the case of $\phi=0.690$, confirming subdiffusive glassy behaviors shown in Fig.~\ref{fig1} (f). 
When $\phi=0.713$ and $\phi=0.753$ in the poly-crystal phase, $\beta(q)$ starts near one and decreases significantly with $q$, which is consistent with the MSD in Fig.~\ref{fig1} (f), verifying again the highly constrained grain-boundary particle motion.
According to Eq.(1) and $F_s(q,t)=exp(-(t/\tau)^{\beta})$, we shall have $\tau(q)\propto q^{-\frac{2}{\beta(q)}}$, which is consistent with the results shown in Fig.~\ref{fig4} (c).  For example, when $\phi=0.111$ and $\phi=0.270$, $\beta(q)\approx-2$ with a weak dependence on $q$, whereas when $\phi=0.713$ $\beta(q)\approx-2$ for small $q$ and approach $-4.8$ at the high $q$ end as shown in Fig.~\ref{fig4} (c). \par

It is curious why the cluster and collective fluids were not observed in the previous experiments \cite{olafsen1998clustering, Shattuck-PhysRevLett.98.188301, Olafsen-Urbach-2005, Pacheco-Vazquez-2009, Aranson-PRL-2000, Howell-PRE-2001, Oyarte-PRE-2013, Rivas-2011a, Rivas-2011b, Rivas-2012, Roeller-PRL-2011, Nahmad-PRE, Neel-PRE-2004, Clewett-2012, Luu-2013}, where a single layer or a shallow layer of spheres is sandwiched between two horizontal plates whose lateral dimension is much larger than their vertical gap and subject to vertical mechanical vibration.  In Ref.~\cite{Aranson-PRL-2000, Howell-PRE-2001, Oyarte-PRE-2013}, electrostatic and magnetic dipole forces introduce long-range interactions that differ substantially from our experiments. The experiments \cite{ Rivas-2011a, Rivas-2011b, Rivas-2012} focus on binary mixtures, which are very different from ours. In Ref.~\cite{Shattuck-PhysRevLett.98.188301, Olafsen-Urbach-2005, Pacheco-Vazquez-2009, Roeller-PRL-2011, Nahmad-PRE, Neel-PRE-2004, Clewett-2012, Luu-2013}, there are only short-range repulsion and inelastic collisions between grains. Despite the quasi-2d characteristics and the similar driving means using a shaker, there are some subtleties between our system and the experiments above.  The main issue is the randomization of particle motion at the single-particle level:  using a flat bottom plate \cite{ olafsen1998clustering, Roeller-PRL-2011, Nahmad-PRE, Neel-PRE-2004, Clewett-2012, Luu-2013} or a cover \cite{guan2021dynamics} introduces a non-Gaussian velocity distribution of a single particle, implying spatial correlations of particle movement, which cannot be eliminated with a rough plate or lid\cite{Olafsen-Urbach-2005, Shattuck-PhysRevLett.98.188301}.  The lack of Gaussian statistics in the single particle could induce phase separations due to a velocity-dependent energy injection rate \cite{Lobkovsky-Urbach-2009, Cafiero-Luding-2000}, as seen in the experiments \cite{olafsen1998clustering, Roeller-PRL-2011, Nahmad-PRE, Neel-PRE-2004, Clewett-2012, Luu-2013}. The phase separation is absent when subject to random forcing, as shown in the simulation\cite{ Lobkovsky-Urbach-2009}.  Before our experiments, two attempts were made to ensure Gaussian statistics of velocities at the single-particle level \cite{Olafsen-Nature, scholz2017velocity}. However, additional effects were introduced, such as the motions of dimers and the continuous particle rotations along a single direction. Therefore, the Gaussian velocity in the single particle driving is crucial. Nonetheless, the direction of the collective particle motion may be related to the asymmetry inherent in the experimental design, which is not the root of the collective motion. Otherwise, the particles would exhibit collective motions at all $\phi$.\par

\paragraph{Conclusion} 
Using pure repulsive Brownian vibrators, we observe various structural and dynamical behaviors, including cluster fluid, collective fluid, poly-crystal, and crystal. In particular, the cluster fluid shows that purely-repulsive hard disks can have effective weak cohesion due to inelastic collision to form particle clusters of intriguing structures and dynamics. A large-scale collective motion emerges near $\phi=0.317$ with the appearance of large clusters and eventually terminates near $\phi=0.713$, close to the melting transition of the equilibrium hard disks. 
Our investigation provides direct microscopic evidence that granular materials subject to uniform random forcing are weakly cohesive with complex internal structures. Moreover, our experiment shows that large-scale collective motion can arise in a purely repulsive hard-disk system.

\begin{acknowledgments}
Y.C. and J.Z. acknowledge the NSFC (No. 11974238 and No. 12274291) support and the Innovation Program of Shanghai Municipal Education Commission under No. 2021-01-07-00-02-E00138. Y.C. and J.Z. also acknowledge the support from the Shanghai Jiao Tong University Student Innovation Center.
\end{acknowledgments}

\bibliographystyle{apsrev4-2} % Tell BibTeX which bibliography style to use
\bibliography{ref} % Tell BibTeX which .bib file to use (this one is an example file in TexLive's file tree)

\end{document}

\title{Supplementary materials of the ``Clusters and collective motions in Brownian vibrators''}

\author{Yangrui Chen$^1$ and Jie Zhang$^{1,2}$}
\email{jiezhang2012@sjtu.edu.cn}
\affiliation{$^1$School of Physics and Astronomy, Shanghai Jiao Tong University, Shanghai 200240, China}
\affiliation{$^2$Institute of Natural Sciences, Shanghai Jiao Tong University, Shanghai 200240, China}
\date{\today}

\maketitle
In the Supplemental Materials, we describe more details on: (I) Voronoi diagrams at various $\phi's$; (II) the two-dimensional(2d) maps of hexatic order parameter $\psi_6^j$ at various $\phi's$; (III) the global mean hexatic order and the 2d correlations of the hexatic order parameter $\psi_6^j$; (IV) the average curl of the particle displacement field; (V) the probability distribution functions (PDFs) of the particle displacement at different $\phi$ in the cluster fluid, collective fluid, poly-crystal, and crystal; (VI) the radial distribution function $g(r)$ and the comparison between $g(r)$ of the cluster fluid and collective fluid and the corresponding results of the equilibrium hard disks; (VII) the algorithm of cluster identification; (VIII) the fitting parameters of the cluster size distributions.

\section{\label{sec:level1} Voronoi diagrams} 
\begin{figure*}[htpb]
\centering
\includegraphics[width=17.2cm]{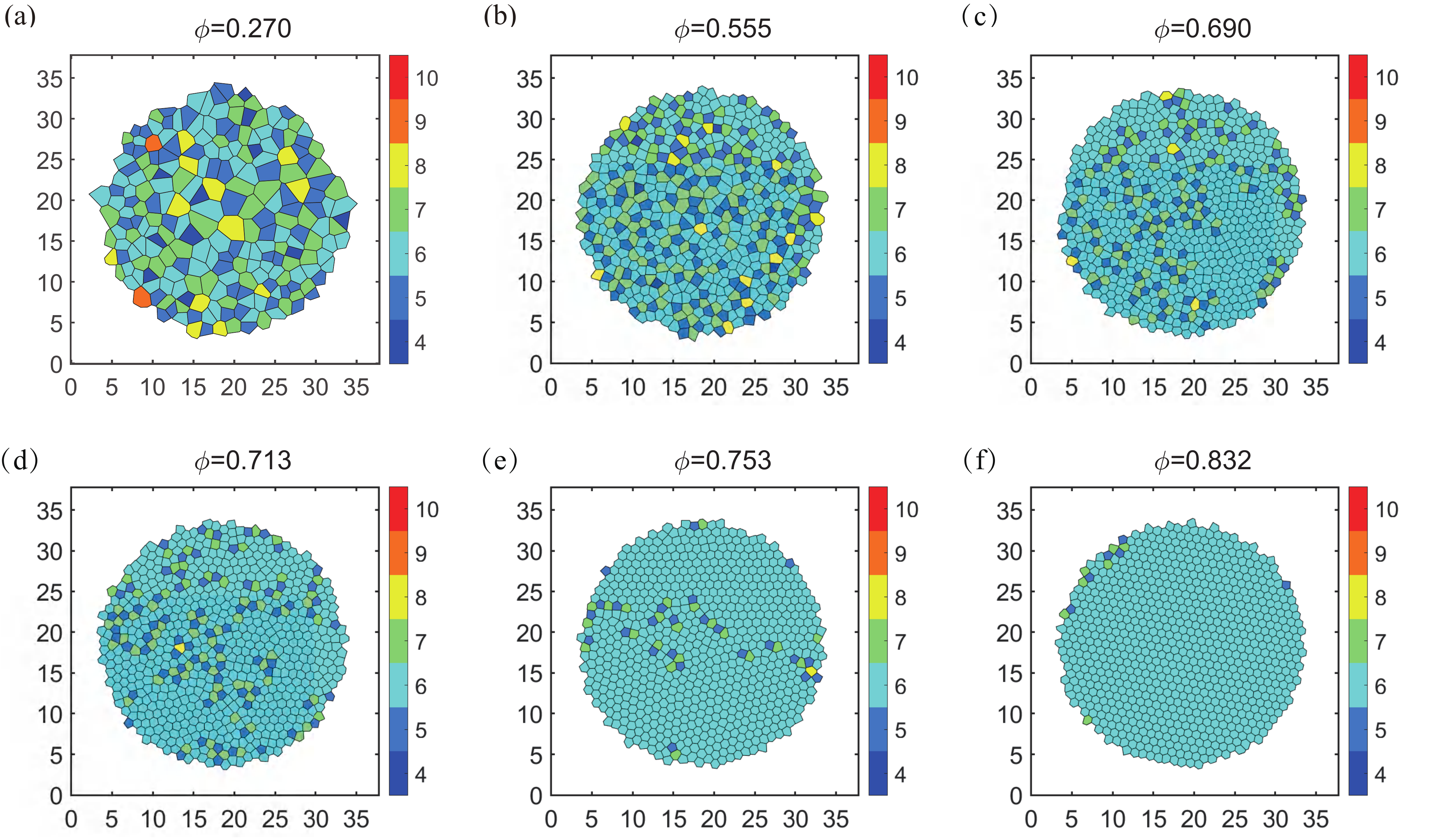}
\caption{\label{smfig1} (a-f) Voronoi diagrams at various $\phi's$.
}
\end{figure*}
Fig.~\ref{smfig1} shows the Voronoi diagrams at different $\phi’s$. In the cluster fluid phase at $\phi=0.270$, the order of a Voronoi cell, i.e., the edge number of the Voronoi cell, ranges from 4 to 9 and is randomly distributed in space. In the collective fluid phase at $\phi=0.555$, the order ranges from 4 to 8, and there is no significant localized ortho-hexagonal aggregation. As a comparison, in the poly-crystalline phase at $\phi_p=0.713$, most Voronoi cells are of order six and with the grain boundaries consisting of pairs of grains of orders 5 and 7. A similar is true for $\phi=0.753$. Except for the defects near boundaries, the crystalline region of $\phi=0.832$ consists of Voronoi cells of order six that spread over the whole system.

\section{\label{sec:level1} The hexatic order parameter} 
\begin{figure*}[htpb]
\centering
\includegraphics[width=17.2cm]{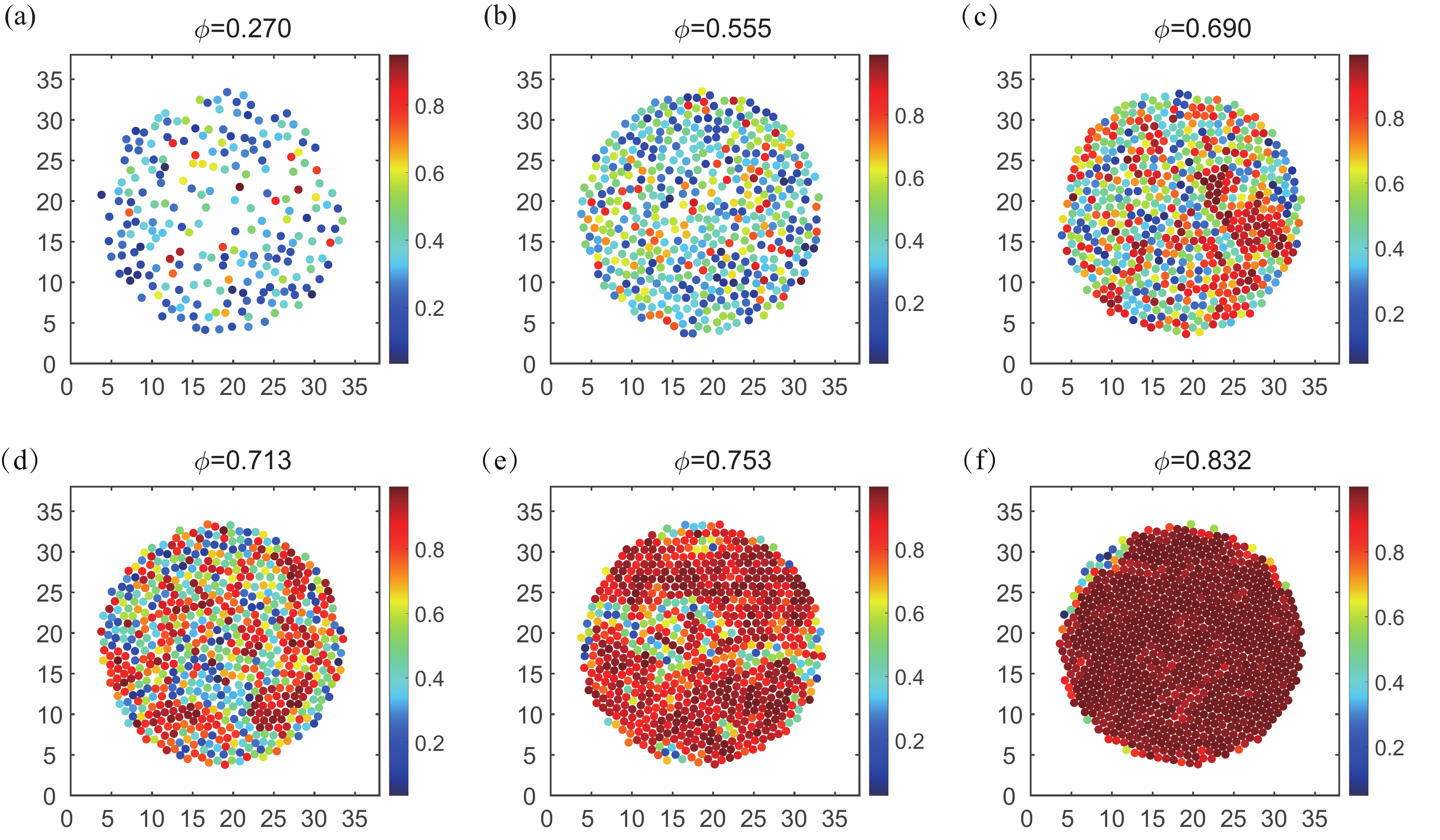}
\caption{\label{smfig2} (a-f) The hexatic order parameter $\psi_6^j$ at various $\phi's$.
}
\end{figure*}
The 2d maps of the hexatic order parameter $\psi_6^j$ at $\phi’s$ are shown in Fig.~\ref{smfig2}, where each panel corresponds to the same particle configurations in Fig.~\ref{smfig1}. The hexatic order parameter $ \psi_6^j=\frac{1}{n_j}\sum_{k=1}^{n_j}e^{i6\theta_{jk}}$ characterizes the similarity of the structure near a particle $j$ with the perfect hexagonal arrangement. Here, $n_j$ is the number of nearest neighbors of the particle $j$ and $\theta_{jk}$ is the angle between the vector $\vec r_{jk}\equiv\vec r_k- \vec r_j$ and the x-axis. Note that $n_j$ does not necessarily equal six: $n_j=2$ describes a linear structure, $n_j=3$ describes a trident structure, $n_j=4$ means a center particle with four neighbors, and et al.  
Let us take $n_j=2$ as an example; when the center particle and its two neighbors form a straight line, $\psi_6^j=1$. Thus, $\psi_6^j$ characterizes local particle-scale symmetry.

\section{\label{sec:level1} The 2d correlations of the hexatic order parameter} 
\begin{figure}[htpb]
\centering
\includegraphics[width=8.6cm]{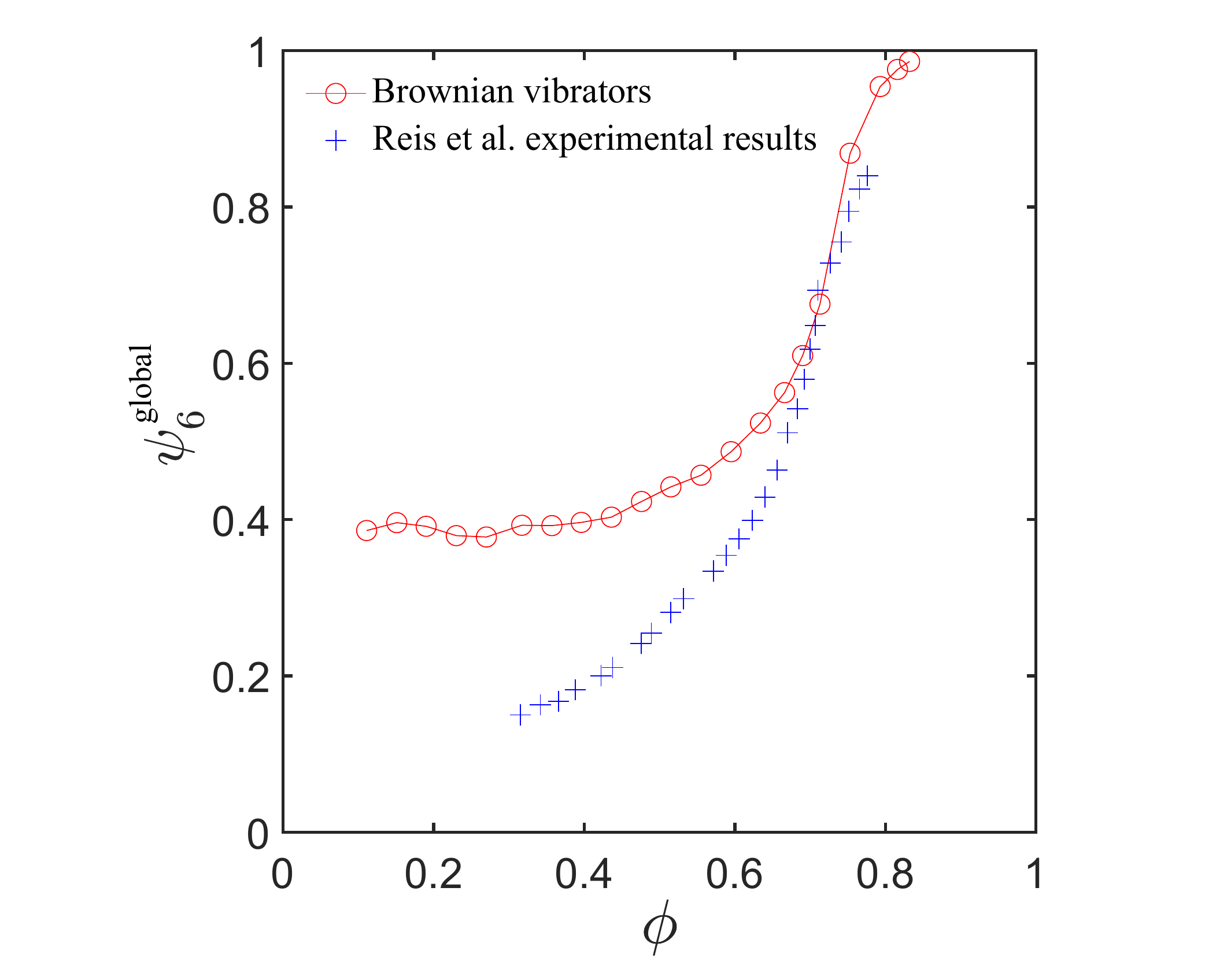}
\caption{\label{smfig3} The global mean of $\psi_6$ at various $\phi's$. Red circles refer to Brownian vibrators. Blue pluses are the results of Ref.\cite{shattuck-PRL-crystal}.
}
\end{figure}

\begin{figure*}[htpb]
\centering
\includegraphics[width=17.2cm]{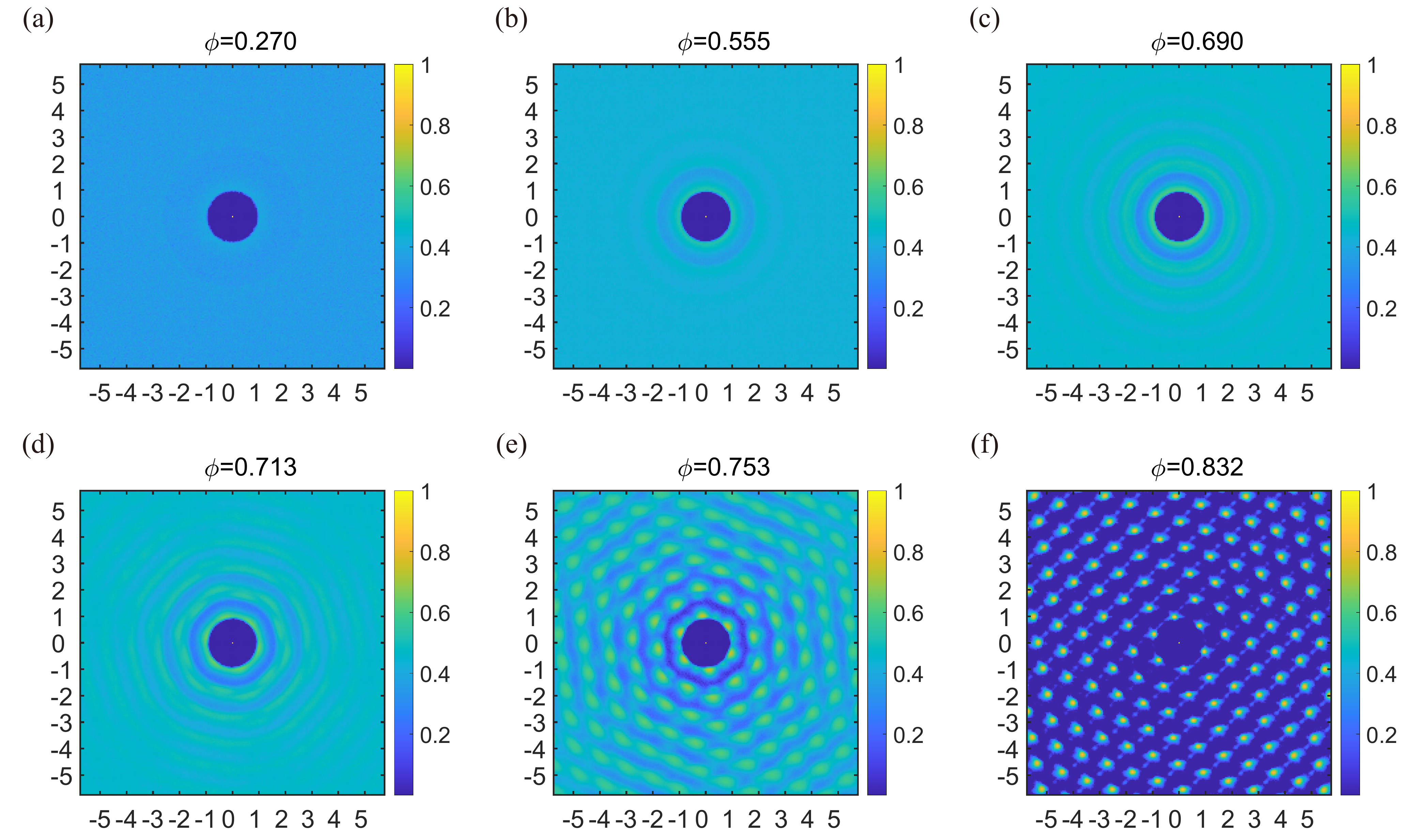}
\caption{\label{smfig4} (a-f) The 2d correlations of the local hexatic order $g^{2D}_6(r,\theta)$ at various $\phi's$.
}
\end{figure*}

The global mean $\psi_6^{global}= \langle \vert \frac{1}{N} \sum_{j=1}^N \psi_6^j \vert \rangle$, where N is the total number of the particles and $\langle ... \rangle$ represents the average over different particle configurations. 
Here, we compare $\psi_6^{global}$ in our experiment with those of Ref.\cite{shattuck-PRL-crystal} in Fig.~\ref{smfig3}, which shows that our system is locally more ordered than that of  Ref.\cite{shattuck-PRL-crystal}, especially leading a plateau for packing fractions $\phi \le 0.436$. 
\par

The two-dimensional correlations of the local hexatic order $g^{2D}_6(r,\theta)= \frac{L^2}{r\Delta r \Delta \theta N(N-1)} \sum\limits_{j\neq k} \delta(r-|\vec{r} _{jk}|) \delta(\theta-|\theta_{jk}|) \psi^j_6 \psi^k_6 $ are shown in Fig.~\ref{smfig4}. 
It is clear that $g^{2D}_6(r,\theta)$ are isotropic for $\phi \le 0.690$ while they show hexagonal symmetry for $\phi\ge0.713$, showing clear crystalline characteristics.  \par

\clearpage
\section{\label{sec:level1} The average curl of the particle displacement field}
We focus on the particle displacement field to quantify the large-scale collective motion.
First, the particle density field can be coarse-grained with the kernel function $\Psi(\vec r)=e^{-(2\vec r/D)^2}$:
\begin{equation}\label{eq1}
\rho(\vec r)=\sum_{i=1}^N \Psi(\vec r-\vec r_i). 
\end{equation} 
Next, the particle displacement field can be coarse-grained within a time interval $\Delta t$:
\begin{equation}\label{eq2}
\vec u(\vec r,\Delta t)=\rho(\vec r)^{-1} \sum_{i=1}^N \vec u_i(\Delta t) \Psi(\vec r).
\end{equation}
Here, the displacement of the particle $i$ is $\vec u_i(\Delta t)=(\vec r_i(t_0+\Delta t)-\vec r_i(t_0))$, and the collective behavior is evident in Fig. 1 (e) in the main text for $\Delta t=100s$.
In a quasi-two-dimensional system, the average curl of the displacement field $\Omega$ is defined as:
\begin{equation}\label{eq3}
\begin{split}
\Omega&=\langle \left[ \nabla \times \vec u(\vec r)\right]_z \rangle \\
&=\langle \rho(\vec r)^{-2} \sum_{i=1}^N\sum_{j=1}^N \Psi(\vec r)_j \left[ \nabla \Psi(\vec r)_i \times \vec u_{ij} \right]_z \rangle.
\end{split}
\end{equation}
Here, the angle brackets represent the spatial average followed by a time average over different $t_0$. 

\clearpage
\section{\label{sec:level1} The PDFs of particle displacement} 

\begin{figure*}[htpb]
\centering
\includegraphics[width=17.2cm]{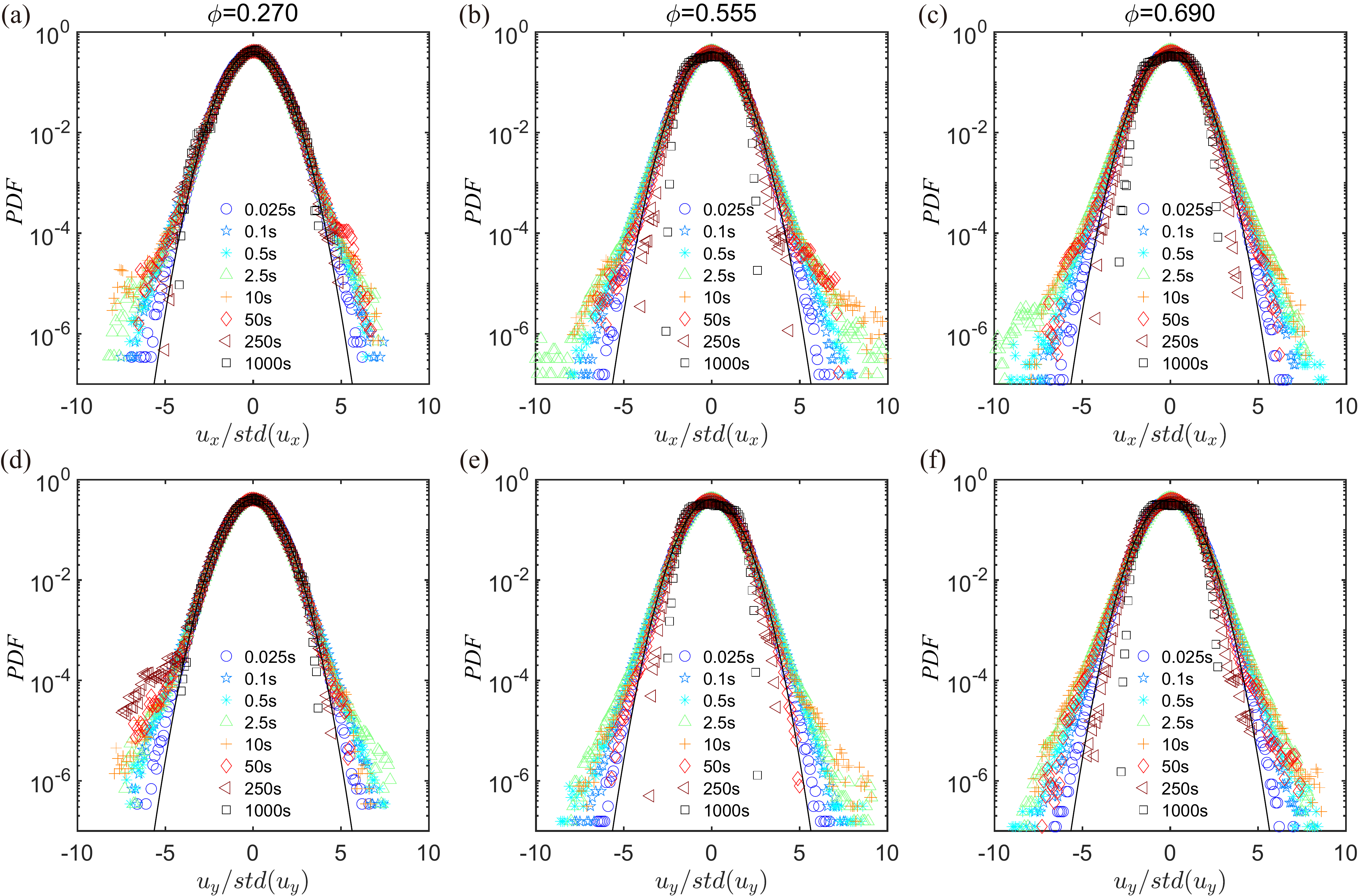}
\caption{\label{smfig5} The PDFs of $u_{\alpha}/std(u_{\alpha})$ ($\alpha$ refers to the $x$ or $y$ component) at the volume fractions of $\phi=0.270,0.555,0.690$. Different colors and markers represent various time intervals. The black solid lines refer to Gaussian distributions. (a-c) $u_x/std(u_x)$ (d-f)$u_y/std(u_y)$.}
\end{figure*}

\begin{figure*}[htpb]
\centering
\includegraphics[width=17.2cm]{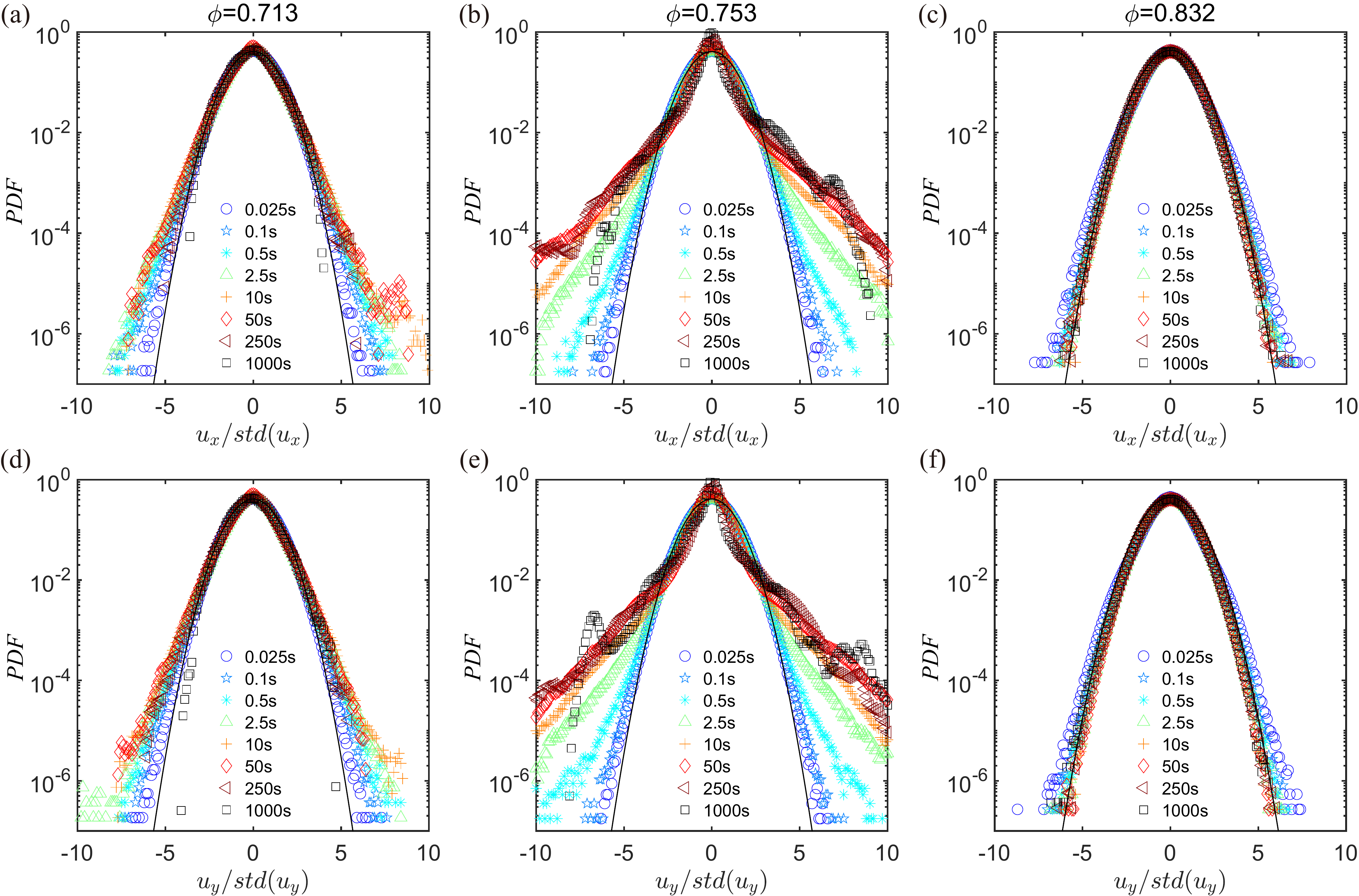}
\caption{\label{smfig6} The PDFs of $u_{\alpha}/std(u_{\alpha})$ ($\alpha$ refers to the $x$ or $y$ component) at the volume fractions of $\phi=0.713,0.753,0.832$. Different colors and markers represent various time intervals. The black solid lines refer to Gaussian distributions. (a-c) $u_x/std(u_x)$ (d-f)$u_y/std(u_y)$.}
\end{figure*}

The PDFs of $u_x$ and $u_y$(the x and y components of the particle displacement $\vec u$) are shown in Fig.~\ref{smfig5} and Fig.~\ref{smfig6}, respectively, where we normalize the horizontal coordinates with the corresponding standard deviation of particle displacement. The solid black lines are Gaussian distributions for comparison. 
None of the PDF curves are Gaussian: each curve departs progressively away from the Gaussian distribution as $u_x/std(u_x)$ or $u_y/std(u_y)$ varies above and below zero. However, in most cases except at $\phi=0.753$, the Gaussian distributions can provide at least an approximate description of the PDFs except in the tails of PDFs with probabilities less than $10^{-3}$.\par

At long time intervals, the PDFs of the distributions of $u_x/std(u_x)$ and $u_y/std(u_y)$ at volume fraction $\phi=0.753$ are most significantly different from Gaussian due to the coexistence of a large number of nearly perfect crystal particles and a relatively small number of grain boundary particles.

\begin{figure*}[htpb]
\centering
\includegraphics[width=17.2cm]{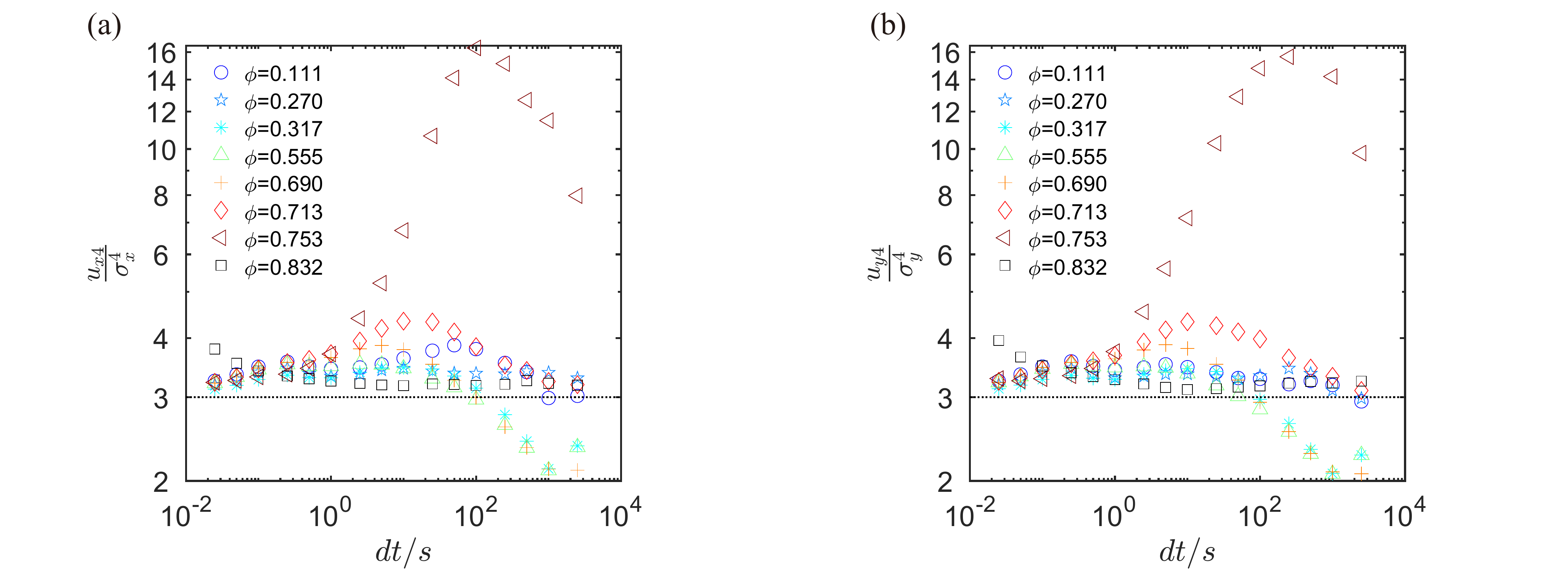}
\caption{\label{smfig7} The kurtosis of the PDFs of $u_x/std(u_x)$ (a) and $u_y/std(u_y)$ (b) for a series of time intervals and different packing fractions $\phi's$. }
\end{figure*}

The quantitative departure of the PDFs from Gaussian can be captured using the kurtosis in Fig.~\ref{smfig7}, which is defined as:  
\begin{equation}\label{eq4}
\frac{\mu_4}{\sigma^4}=\frac{\frac{1}{n}\sum_{i=1}^n(u_i-\bar{u})^4}{(\frac{1}{n}\sum_{i=1}^n(u_i-\bar{u})^2)^2}.
\end{equation}
Note that the kurtosis of Gaussian equals 3. Here $u_i$ is a random variable, which may refer to either $u_x/std(u_x)$ or $u_y/std(u_y)$.\par

\clearpage

\section{\label{sec:level1} The radial distribution function} 

The radial distribution function, also known as the pair correlation function, is defined as: 
\begin{equation}\label{eq5}
g(r)= \frac{L^2}{2 \pi r \Delta r N(N-1)} \sum\limits_{j\neq k} \delta(r-|\vec{r}_{jk}|) ,
\end{equation}
where $L$ is the system size, $N$ is the total number of particles, $|\vec{r}_{jk}|$ refers to the relative distance between two particles $j$ and $k$, and $\Delta r$ is a small increment along the radial direction.
\par

\begin{figure*}[htpb]
\centering
\includegraphics[width=17.2cm]{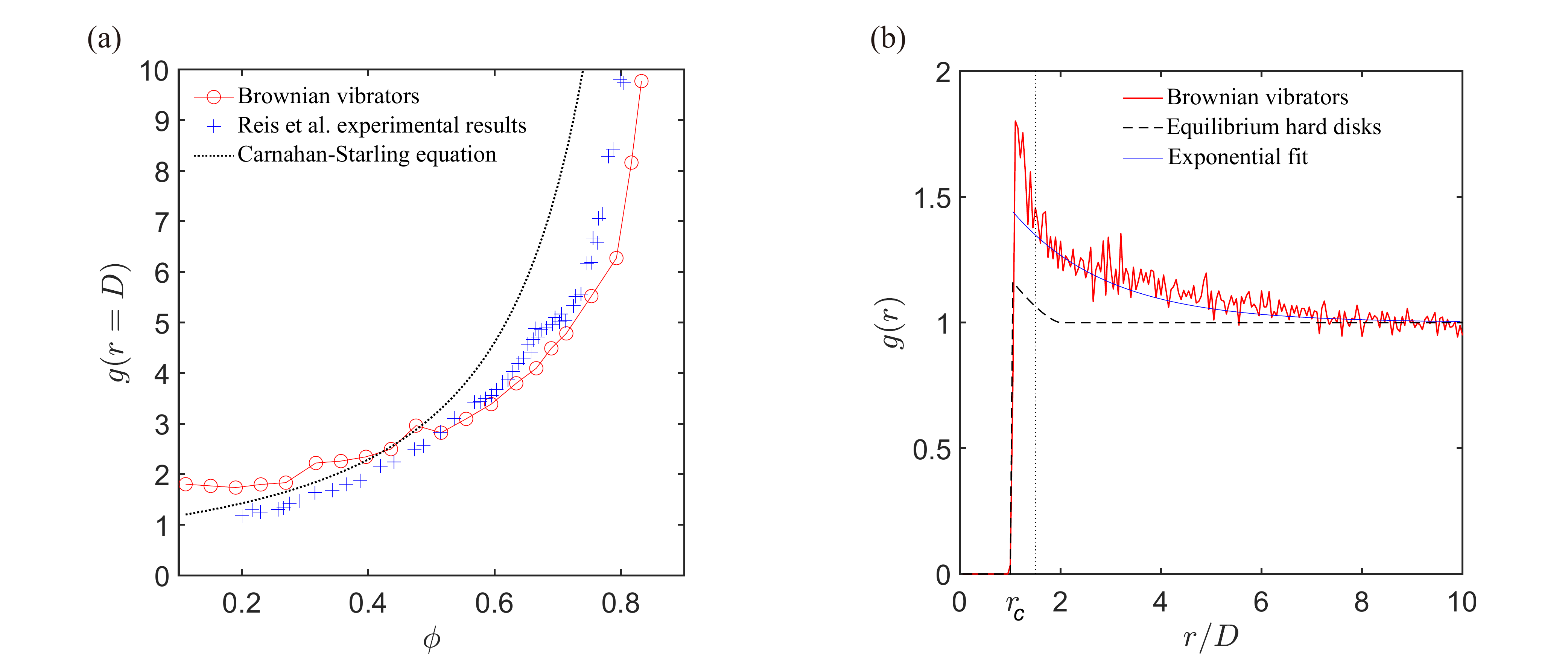}
\caption{\label{smfig8} (a) $g(r = D)$ versus $\phi$. Red circles refer to our experimental results of Brownian vibrators. Blue pluses are the experimental results with vibrated stainless steel spheres in Ref.\cite{shattuck-PRL-crystal}. The black dotted line is the Carnahan-Starling equation. (b) The red line shows $g(r)$ at $\phi=0.111$ of our Brownian vibrators. The blue line is an exponential fit for $r\ge 1.5D$. The black dashed line is the theoretical curve of the equilibrium hard disks \cite{chae1969radial}.
}
\end{figure*}

The particle configurations of our system are different from those of equilibrium hard disks and the previous experiment using vibrated stainless steel spheres Ref.\cite{shattuck-PRL-crystal}, as can be seen from the curve of 
$g(r=D)(\phi)\equiv g_{D}(\phi)$ shown in Fig.~\ref{smfig8} (a) with $D$ being the particle diameter. The black dotted line refers to the theoretical curve of Carnahan-Starling equation $g^{CS}_{D}(\phi)=[1-7/16\phi]/(1-\phi)^2$, which is derived from the equation of state for non-attracting hard spheres\cite{carnahan1969equation}. \par

In Fig.~\ref{smfig8} (b), we plot the $g(r)$ at $\phi=0.111$, from which we define the threshold $r_c$, corresponding to the half-height of the first peak. The blue solid line shows that $g(r)$ decays exponentially for $r>r_c$ as $g(r)\propto e^{-(r-1)/(\xi-1)}$, whereas it decays more rapidly for $r\le r_c$, with $\xi \approx 2.9\pm 0.4$. Note that the peak is much higher than that of the equilibrium hard disks\cite{chae1969radial}.

\newpage

\section{\label{sec:level1} The algorithm of identifying a cluster} 

We define a cluster as a set of percolated particles with the distance between the percolated two neighboring particles smaller or equal to $r_c$ such that any particle outside the cluster is at a distance greater than $r_c$ from all particles in the cluster. Using $r_c$, starting from a given particle $A$, we run a search to find all possible percolated neighboring particles $B$ whose centers are within the radius $r_c$ from $A’s$ center. If there is no $B$, $A$ is a single-particle cluster. If there are multiple $B’s$, for each $B$ particle, we search its percolated neighbors, and we run the algorithm recursively till finding no more percolated particles.

\section{\label{sec:level1} The fitting parameters of the cluster size distributions} 

\begin{figure*}[htpb]
\centering
\includegraphics[width=17.2cm]{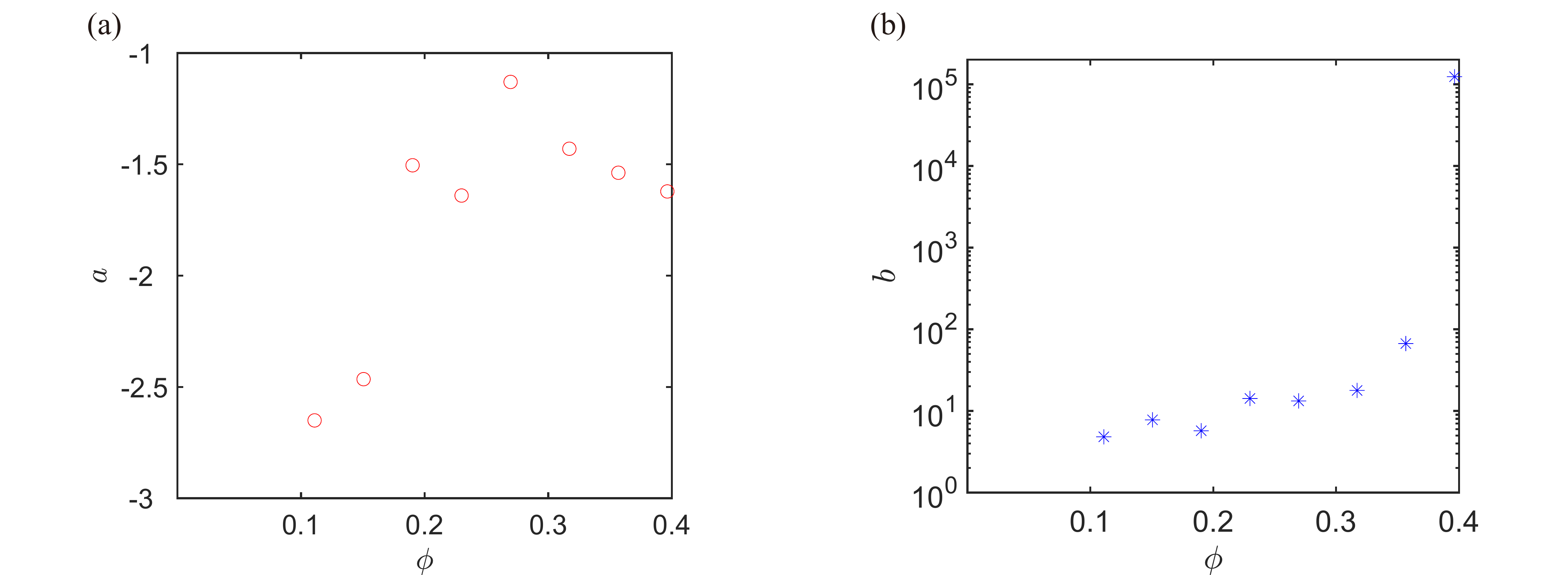}
\caption{\label{smfig9} (a-b) The fitting parameters $a$ and $b$ of $PDF \propto N^{a}e^{-N/b}$ for the distributions of the cluster size $N$ at different $\phi's$. 
}
\end{figure*}

In the fit of the PDFs of the cluster size $N$, we assume a power-law distribution with an exponential cutoff, i.e., $PDF \propto N^{a}e^{-N/b}$. The fitting parameters $a$ and $b$ are shown in the above figures, where $a$ peaks at $\phi=0.270$ and gradually decreases while entering the collective fluid phase. The fitting parameter $b$ diverges at $\phi \ge 0.396$, indicating that the distribution fits quite nicely with a power-law distribution, and a correction with an exponential cutoff is no longer necessary.

\bibliographystyle{apsrev4-2} % Tell BibTeX which bibliography style to use
\bibliography{sref} % Tell BibTeX which .bib file to use (this one is an example file in TexLive's file tree)